# Magneto-transport and domain wall scattering in epitaxy $L1_0$ MnAl thin film

Linqiang Luo[1], Nattawut Anuniwat[1], Nam Dao[2], Yishen Cui[1], Stuart A. Wolf[1,2,3], and Jiwei Lu[2]

[1] Department of Physics, University of Virginia, Charlottesville, Virginia, 22904, USA

[2] Department of Materials Science and Engineering, University of Virginia, Charlottesville, Virginia, 22904, USA

[3] Institute of Defense Analyses, Alexandria, Virginia, 22311, USA

**Abstract**

Epitaxial L1$_0$ MnAl films demonstrated two different kinds of magneto-transport behaviors as a function of temperature. The magneto-resistance ratio (MR) was negative and exhibited evident enhancement in the resistivity at coercive fields above ~175 K. The MR enhancement was attributed to the increase of the magnetic domain walls based on the quantitative correlation between the domain density and the resistivity. Below 175 K, the MR was positive and showed a quadratic dependence on the external magnetic field, which implied that the MR was dominated by Lorentz effects below 175 K.

**Introduction**

Permanent magnetic thin films with perpendicular magnetic anisotropy (PMA) have been of great interest due to their application in perpendicular magnetic recording and spintronic devices.[1,2] Among the L1$_0$ intermetallics, MnAl shows attractive magnetic properties.[3,4] The saturation magnetization of 490 emu/cm$^3$ along with large perpendicular uniaxial magnetocrystalline anisotropy $K_U$ ~10$^7$ erg/cm$^3$ was obtained in bulk MnAl.[5] The high $K_U$ makes it promising as a magnetic fixed layer for perpendicular magnetic tunnel junctions (p-MTJs).[6] The low damping constant of MnAl (~0.006)[6] is very desirable for low energy dissipation spin torque devices such as spin-transfer torque magnetic tunnel junction and spin torque nano-oscillators (STNO).[7] For example, STNO operation originally required very large external magnetic fields, recent innovation in utilizing PMA materials has made low to zero field operation possible.[8-12]

The effect of magnetic domain walls (DWs) on the magneto-transport behavior of thin films is a topic of great interest for fundamental studies and for domain wall motion related applications.[13-15] However, the measurement of DWs' resistance remains challenging because the resistance from DWs scattering is very small and can be comparable with the anisotropic magneto-resistance (AMR). Epitaxial films with strong perpendicular magnetic anisotropy, such as hcp(0001) Co film,[16,17] $L1_0$ FePt[18] and FePd,[19] are ideal systems for the study of DWs' resistance, since magnetization inside the domains is always perpendicular to the current, which excludes the AMR effect.

Here, $L1_0$ MnAl's magneto-transport properties as a function of temperature have been investigated using a Hall bar structure. The magnetic domain structures were characterized by magnetic force microscope (MFM). We analyzed the impact of domain wall scattering on magneto-resistance.

**Experiment**

MnAl films were synthesized by using co-sputtering of Mn and Al targets in a Biased Target Ion Beam Deposition system (BTIBD). The deposition details can be found elsewhere.[20] The complete structure was MgO (substrate) / 40 nm Cr / 20 nm MnAl / 10 nm Cr. The 40 nm Cr seed layer was used to reduce the lattice mismatch between the substrate and the MnAl. The capping layer was used to prevent MnAl from oxidization.

The film thickness and phase composition were characterized using a high resolution X-ray diffractometer with Cu Kα radiation (Smart-lab, Rigaku Inc.). The in-plane and out-of-plane magnetic hysteresis loops were measured from 50 K to 320 K by a vibrating sample magnetometer (PPMS 6000, Quantum Design). The magnetic domain structure was studied by a magnetic force microscope (Cypher, Asylum Research) with a CoCr coated Si tip at room temperature.

To measure the magneto- and Hall resistances, the films were patterned into a Hall bar using two-step photolithography and ion milling. The Hall bar had a 50μm linewidth with contacts for measurement of both longitudinal and transverse (Hall) resistance. Longitudinal magneto-resistance ($R_{xx}$) measurements were performed on the Hall bar in the temperature range 50 K-320 K and in magnetic fields up to 30 kOe, with the field applied: (1) in the film plane and parallel to the electrical current [longitudinal configuration], (2) perpendicular to the film plane [perpendicular configuration]. The difference of spatial geometry was achieved by manually rotating the sample with respect to the field direction.

**Results and discussion**

The structural properties of MnAl films were characterized by the XRD 2θ scan as shown in Fig. 1(a). In addition to the diffraction peaks from the Cr seed layer and MgO substrate, two diffraction peaks from MnAl were observed at 25.21 ° and 51.59 °, corresponding to the fundamental (002) peak and superlattice (001) peak respectively. The superlattice peak indicated the presence of a tetragonal distorted $B_2$ structure of MnAl, thanks to the chemical ordering of Mn and Al. The order parameter $S$ for the MnAl was estimated to be ~0.89 using integrated intensities of the (001) and (002) peaks.[21] This indicated a very high degree of chemical ordering in the MnAl thin film ($S$=1 represents perfect chemical ordering).

The room temperature M-H loops of the same sample are shown in Fig. 1(b). The result showed a clear indication of perpendicular magnetic anisotropy. The saturation magnetization ($M_S$) was calculated to be 462.5 emu/cm$^3$, and the magnetic anisotropy energy ($K_U$) ~5.3×10$^6$ erg/cm$^3$. Both values were comparable to those of the bulk.[5] It is worth mentioning that these values were considerably higher than the corresponding values for MnAl films directly deposited on a MgO substrate.[20,22] The Cr seed layer had a much smaller lattice mismatch with the MnAl film as compared to MgO, which led to a larger tetragonality in MnAl with a c/a~1.26, while c/a was ~1.0 for MnAl that was grown directly on the MgO substrate.[20] As predicted,[23,24] the c/a of 1.26 resulted in a higher magnetic moment.

The magnetic domains of the MnAl films were characterized by MFM as shown in Fig. 1(c). Maze-like stripe domains were observed in the MFM phase image, which was expected for magnetic thin films with strong PMA. The magnetic domain structure is strongly dependent on the ratio of anisotropy to magneto-static energy: $Q=K_U/2\pi M_S^2$, where $K_U$ is the uniaxial anisotropy constant, $M_S$ the saturation magnetization. $Q$ was ~4 based on the values of $K_U$ and $M_S$ from the VSM results at room temperature. For $Q>1$, stripe domains that intersect the surface with PMA are energetically favored.[19,25]

Figure 2 shows the isothermal magnetoresistance ratio (MR) (defined as $\Delta R/R=[R(H,T)-R(0,T)]/R(0,T)$) versus field curves measured from 50 K to 320 K; the external field was applied out of plane. Above 175 K, the sign of the MR was negative, and an evident two-peak shape was observed. A sign change in MR occurred near 175 K, at which temperature the co-existence of both a negative and positive MR was apparent. Below 175 K, The sign of MR was positive and increased monotonically when the temperature was reduced.

The Hall resistivity ($\rho_{xy}$) of the MnAl Hall bar was measured from 175K to 320K. Fig. 3(a) inset shows the Hall resistivity ($\rho_{xy}$) as a function of field at 300K. According to Pugh's equation given by[26]

$$\rho_{xy} = R_0 H_z + R_s M_z \quad (1)$$

where $H_z$ is the perpendicular field, $M_z$ is the magnetization, $R_0$ and $R_s$ are the ordinary and anomalous Hall coefficients. The first linear component comes from the normal Hall Effect, while the second extraordinary component is ascribed to the anomalous Hall Effect (AHE). The Hall resistivity was hysteretic, mimicking the magnetic hysteresis;[27] therefore, the coercivity ($H_C$) can be extracted at each temperature.

From 175 K-320 K, the peak positions were extracted and compared with the $H_C$ obtained from the AHE loop as shown in Fig. 3(a). The MR reached a maximum when the magnetization was close to zero at $\pm H_C$. This implied the existence of magnetic inhomogeneity.[28] In the MnAl films, the inhomogeneity was a consequence of the magnetic domain walls that causes strong spin-dependent scattering on the nano-scale. At the coercive field, more magnetic domains were present thanks to the field-induced domain wall motion and/or domain rotation during the magnetization reversal.[29,30] The increase in the domain density led to the substantial increase in the domain wall induced scattering of charge carriers, hence the enhancement in the resistance.

Anisotropic magnetoresistance (AMR), which contributed to the MR in previous reports,[31] did not have a significant effect on the MR in MnAl, because the angle between the magnetization and current was kept unchanged during the sweeping of the external magnetic field.

Berger proposed that the Hall effect led to a zigzag pathway for the electric current going through stripe domains, and a concomitant increase in the resistivity by $(\rho_{xy}/\rho_{xx})^2$.[32] $(\rho_{xy}/\rho_{xx})^2$ was calculated based on the Hall measurements. The results were then compared with the MR enhancement at the coercive field from 175 K to 320 K as shown in shown in Fig. 3(b). The MR enhancement (defined as $[R(H_C)-R(0)]/R(0)$) at coercivity of 0.013% was one order of magnitude larger than $(\rho_{xy}/\rho_{xx})^2$ (~0.001%) at 175 K. Thus, the Hall effect contribution was of insufficient magnitude to explain the MR enhancement observed here.

Resistivity measurements with the external field applied in the longitudinal direction ($\rho_{long}$) and the perpendicular direction ($\rho_{perp}$) are shown in Fig. 4(a), where $\rho_{long}$ is larger than $\rho_{perp}$. MFM images of the same film were scanned in an area of 100 μm$^2$. Fig. 4(b) and Fig. 4(c) show the measured MFM images corresponding to the remanent states of longitudinal and perpendicular measurements, which are marked as points A and B in Fig. 4(a) respectively. The MFM image demonstrated a larger domain size in B, which can be ascribed to the result of the magnetization process.[33] After removing the out-of-plane field, the uniform magnetized domain broke up into multiple domains to minimize the demagnetizing energy. The magnetic moments in B had to overcome the uniaxial anisotropy and rotated toward the opposite direction to form a domain wall. In the case of A, since the magnetic field already aligned the magnetic moments to an in-plane direction, the magnetic moments could either rotate up or down with no preferred orientation, and therefore it was much easier to form magnetic domains in A. By defining the boundary of domains in the MFM images using the software WSxM,[34] the number of domains can be counted statistically and the domain density has been calculated, which gives the domain density of $5.57/\mu m^2$ and $4.67/\mu m^2$ for case A and case B, respectively. The in-plane field leads to a higher domain density hence a higher density of DWs, which give rise to the resistivity.

In order to extract the DWs scattering contribution to the resistivity quantitatively, the resistivity is expressed by[19]

$$\rho_{long} = \rho_s + \frac{\delta_w}{d_{long}}\rho_{DW} + \rho_{AMR//} \approx \rho_s + \frac{\delta_w}{d_{long}}\rho_{DW} \quad (2)$$

$$\text{and } \rho_{perp} = \rho_s + \frac{\delta_w}{d_{perp}}\rho_{DW} + \rho_{AMR\perp} \approx \rho_s + \frac{\delta_w}{d_{perp}}\rho_{DW} \quad (3)$$

where $\rho_{long}$ and $\rho_{perp}$ are the measured resistivity, $\rho_s$ the resistivity in saturated state, $\rho_{DW}$ the extra domain wall induced resistivity, $\rho_{AMR//}$ and $\rho_{AMR\perp}$ the AMR contributions corresponding to case A and B. At the low magnetic field, the MFM images revealed an out-of-plane magnetization for both cases, so the $\rho_{AMR}$ is estimated to be much smaller than $\rho_s$. $d_{long}$ and $d_{perp}$ are the average domain sizes, which were estimated from the MFM images, 200 nm and 275 nm for A and B respectively. $\delta_w$ is the domain wall width. For films with $Q>1$, the domain walls are assumed to be Bloch Walls in the center of the film[35]. The wall width is given by equation[33]

$$\delta_{Bloch} = \pi\sqrt{A/K_U} \quad (4)$$

where $A$ is the exchange stiffness constant and $K_U$ the uniaxial anisotropy constant $5.3\times10^6$ erg/cm$^3$. The exchange stiffness for L1$_0$ MnAl has been estimated by[36]

$$A = 3k_BT_C/2za \quad (5)$$

where $k_B$ is the Boltzmann constant ($1.38\times10^{-16}$ erg/K), $T_C$ is the Curie temperature (690K),[37] $z=8$ is the number of nearest neighbor atoms, and $a$ is the lattice constant (2.80Å). Therefore, $A= 6.38\times10^{-7}$ erg/cm. The wall width for MnAl was then calculated to be $\delta_w \sim 10.9$ nm, which was comparable to the values found in FePt ($\delta_w=6$ nm)[38] and CoCrPt ($\delta_w=14$ nm)[36]. By combining Eqn. (2) and Eqn. (3), and substituting in the above-mentioned information, the equations can be solved in terms of $\rho_{DW}/\rho_s$. The DWs scattering contribution to the resistivity is calculated to be $\rho_{DW}/\rho_s= 1.75\%$ for the maze states as shown in Fig. 4(b)&(c). This value was smaller than the current-perpendicular-to-wall (CPW) (8.2%) DWs contribution but larger than the current-in-wall (CIW) (1.3%) DWs contribution for FePd that had parallel stripe domains.[19] Unlike FePd, MnAl had a maze-like stripe domain that possibly results in an average of current conduction channels between CPW and CIW geometry. Based on the averaging of the conductions along the two orientations, the value was in good agreement with the theoretical prediction of Levy and Zhang for typical ferromagnetic materials.[14]

Below 150K, the MR became positive, and was quadratic with respect to the external magnetic field, which was attributable to the Lorentz force on the trajectories of the electrons. The Lorentz force leads to an orbital motion for electrons in an out-of-plane magnetic field, and the Lorentz MR is predicted to increase quadratically as a function of the B field. [39,40]

The MR measured with the field direction parallel to the current direction at 50 K and 250 K are shown in Fig. 5. The sign of the MR was negative for both temperatures and no quadratic relation was observed. Here the Lorentz effect was minimized due to the parallel alignment between the electric current and magnetic field. This confirms the fact that the positive MR at low temperatures was caused by the extrinsic contribution from the Lorentz effect. At 50 K, at high field, the MR showed an upturn to become positive, which may be due to the small misalignment (<5°) of the hall bar with the field direction. It is noted that the Lorentz MR can be suppressed due to the decrease of mean scattering time when the temperature increases,[40] which may explain the sign change in the MR at ~ 175 K.

**Conclusions**

In summary, magneto-transport properties of $L1_0$ MnAl films with high chemical ordering ~0.89 and strong PMA ~$5.3 \times 10^6$ erg/cm$^3$ has been investigated. Maze-like stripe domains with out of plane magnetization were observed by MFM. A temperature-dependent magneto-resistance change was observed on the Hall bar patterned sample. From 320 K to 175 K, the low field MR enhancement was linked to the DWs scattering of charge carriers. Further analysis on remanent states' MFM images and the corresponding resistivity demonstrated the contribution of DWs to the electric resistivity of MnAl. A MR sign change occurred around 175 K, and the MR turned positive and quadratic with respect to field when the temperature was below 175K. The longitudinal measurements suggest that the MR from Lorentz effect became dominant at temperatures below 175 K.

**Acknowledgement**

Authors acknowledge the financial support by the National Science Foundation (Award number: ECCS-1344218). L.L. thanks S.Gider, G.Albuquerque and S.Kittiwatanakul for helpful discussions.**Reference**

1. T. Oikawa, M. Nakamura, H. Uwazumi, T. Shimatsu, H. Muraoka, and Y. Nakamura, Magnetics, IEEE Transactions on **38,** 1976 (2002).
2. S. Mangin, D. Ravelosona, J. A. Katine, M. J. Carey, B. D. Terris, and E. E. Fullerton, Nature Materials **5,** 210 (2006).
3. S. Mizukami, S. Iihama, N. Inami, T. Hiratsuka, G. Kim, H. Naganuma, M. Oogane, and Y. Ando, Applied Physics Letters **98,** 052501 (2011).
4. M. Hosoda, M. Oogane, M. Kubota, T. Kubota, H. Saruyama, S. Iihama, H. Naganuma, and Y. Ando, Journal of Applied Physics **111,** 07A324 (2012).
5. A. J. J. Koch, P. Hokkeling, M. G. v. d. Steeg, and K. J. de Vos, Journal of Applied Physics **31,** S75 (1960).
6. H. Saruyama, M. Oogane, Y. Kurimoto, H. Naganuma, and Y. Ando, Japanese Journal of15

**Figure captions:**

Fig 1. (a) 2θ XRD scan of MnAl showing the fundamental (002) and superlattice (001) peaks. (b) In-plane (Blue triangles) and out-of-plane (Red dots) hysteresis loops for MnAl films measured at room temperature. (c) 2μm × 2μm MFM phase image for the MnAl film after ac demagnetization.

Fig 2. MR curves on a 50 μm wide Hall bar measured at temperature from 50 K to 320 K. The external field was applied out of plane. The inset shows the Hall bar pattern with the longitudinal measurement configuration.

Fig 3. (a) Coercivity (Red dots) versus the MR peak positions (Blue squares) for temperature range 175 K to 320 K. The inset shows the Hall resistivity versus the perpendicular applied field at 300 K. (b) The MR enhancement at coercivity and $(\rho_{xy}/\rho_{xx})^2$ vs temperature.

Fig 4. (a) The resistivity measured with the field applied in the longitudinal direction (Red dots) and the perpendicular direction (Blue triangles), corresponding to $\rho_{long}$ and $\rho_{perp}$ respectively. The remanent states are marked as A and B. (b)&(c) MFM phase images corresponding to remanent states A and B. The domain boundaries have been highlighted in green color. Density of domains in state A: 5.57/μm² and B: 4.67/μm²

Fig 5. MR curves measured at temperature 250 K (Blue dots) and 50 K (Red triangles). The external field was applied in plane and along the longitudinal direction.

*Fig. 1*

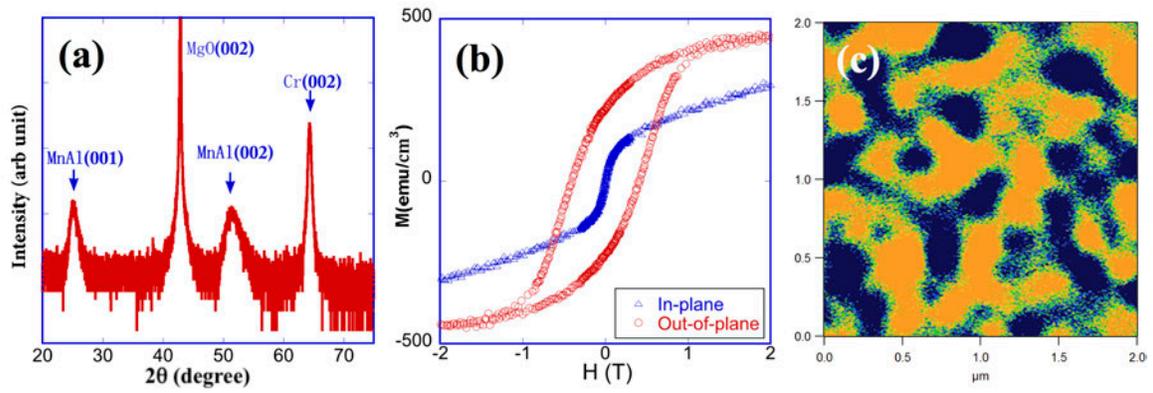

*Fig. 2*

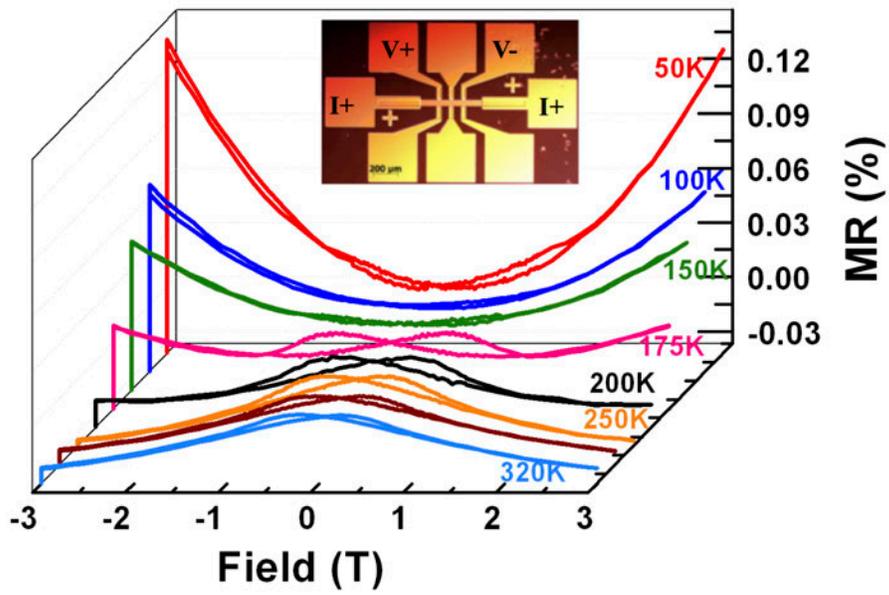

*Fig. 3*

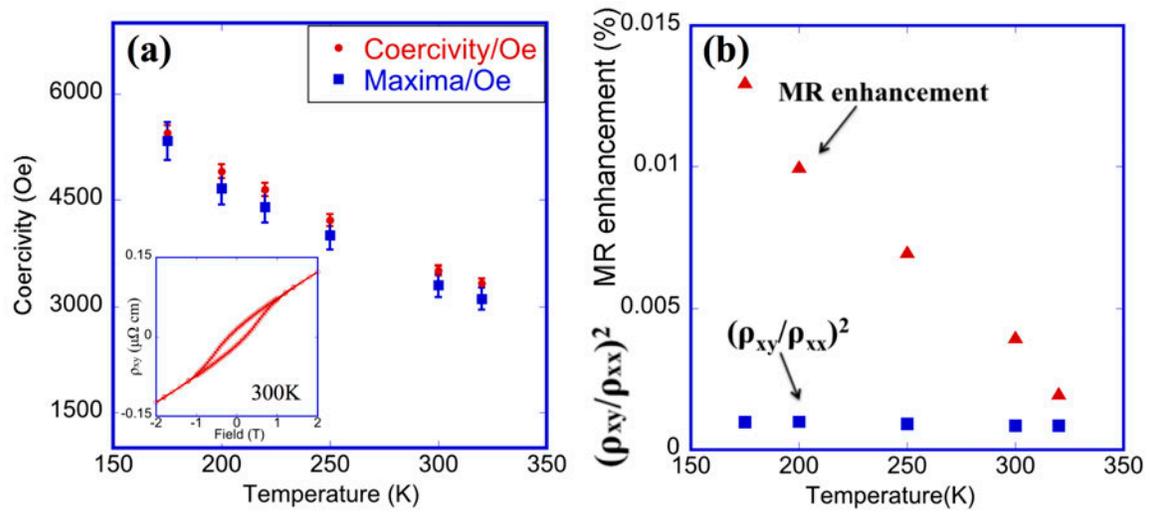

*Fig. 4*

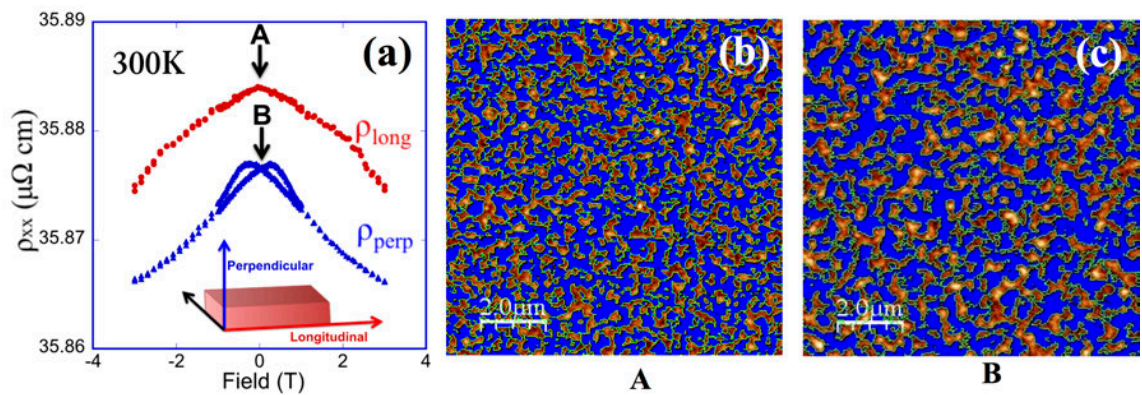

*Fig. 5*

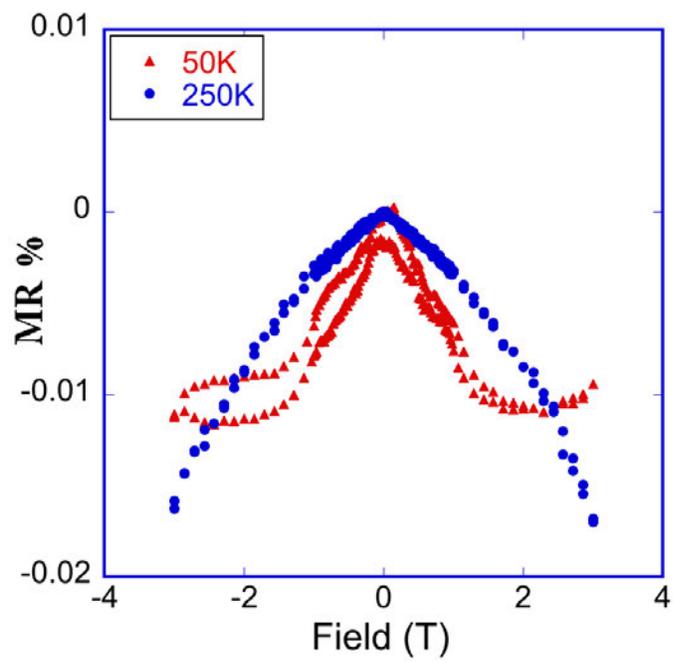